\documentclass[a4paper,UKenglish,cleveref, autoref, thm-restate]{lipics-v2021}
\hideLIPIcs  %uncomment to remove references to LIPIcs series (logo, DOI, ...), e.g. when preparing a pre-final version to be uploaded to arXiv or another public repository

\definecolor{keywordcolor}{rgb}{0.7, 0.1, 0.1}   % red
\definecolor{tacticcolor}{rgb}{0.0, 0.1, 0.6}    % blue
\definecolor{commentcolor}{rgb}{0.4, 0.4, 0.4}   % grey
\definecolor{symbolcolor}{rgb}{0.0, 0.1, 0.6}    % blue
\definecolor{sortcolor}{rgb}{0.1, 0.5, 0.1}      % green
\definecolor{attributecolor}{rgb}{0.7, 0.1, 0.1} % red

\title{Formalizing the Ring of Ad\`eles of a Global Field} %TODO Please add

\author{Mar\'ia In\'es {{de Frutos}-{Fern\'andez}}}{Imperial College London, United Kingdom  \and \url{https://www.imperial.ac.uk/people/m.de-frutos-fernandez}}{m.de-frutos-fernandez@imperial.ac.uk}{https://orcid.org/0000-0002-5085-7446}{}%TODO mandatory, please use full name; only 1 author per \author macro; first two parameters are mandatory, other parameters can be empty. Please provide at least the name of the affiliation and the country. The full address is optional. Use additional curly braces to indicate the correct name splitting when the last name consists of multiple name parts.

\authorrunning{M. I. de Frutos-Fern\'andez} %TODO mandatory. First: Use abbreviated first/middle names. Second (only in severe cases): Use first author plus 'et al.'

\Copyright{Mar\'ia In\'es de Frutos-Fern\'andez} %TODO mandatory, please use full first names. LIPIcs license is "CC-BY";  http://creativecommons.org/licenses/by/3.0/

\ccsdesc[500]{Theory of computation~Logic and verification}
\ccsdesc[500]{Theory of computation~Type theory} %TODO mandatory: Please choose ACM 2012 classifications from https://dl.acm.org/ccs/ccs_flat.cfm 

\keywords{formal math, algebraic number theory, class field theory, Lean, mathlib} %TODO mandatory; please add comma-separated list of keywords

\category{} %optional, e.g. invited paper

\relatedversion{} %optional, e.g. full version hosted on arXiv, HAL, or other respository/website
\funding{EPSRC Grant EP/V048724/1: Digitising the Langlands Program (UK)}%optional, to capture a funding statement, which applies to all authors. Please enter author specific funding statements as fifth argument of the \author macro.

\acknowledgements{I would like to thank Kevin Buzzard for his constant support and for many helpful conversations during the completion of this project, and Ashvni Narayanan for pointing out that the finite ad\`ele ring could be defined for any Dedekind domain. I am also grateful to Patrick Massot for making some of the topological prerequisites available in \texttt{mathlib}, and to Sebastian Monnet for formalizing the topology on the infinite Galois group. Finally, I thank the \texttt{mathlib} community for their helpful advice, and the \texttt{mathlib} maintainers for the insightful reviews of the parts of this project already submitted to the library.}%optional

\nolinenumbers %uncomment to disable line numbering

\EventEditors{John Q. Open and Joan R. Access}
\EventNoEds{2}
\EventLongTitle{42nd Conference on Very Important Topics (CVIT 2016)}
\EventShortTitle{CVIT 2016}
\EventAcronym{CVIT}
\EventYear{2016}
\EventDate{December 24--27, 2016}
\EventLocation{Little Whinging, United Kingdom}
\EventLogo{}
\SeriesVolume{42}
\ArticleNo{23}
\begin{document}

\maketitle

\begin{abstract}
The ring of ad\`eles of a global field and its group of units, the group of id\`eles, are fundamental objects in modern number theory. We discuss a formalization of their definitions in the Lean 3 theorem prover. As a prerequisite, we formalized adic valuations on Dedekind domains. We present some applications, including the statement of the main theorem of global class field theory and a proof that the ideal class group of a number field is isomorphic to an explicit quotient of its id\`ele class group.
\end{abstract}

\section{Introduction}
\label{sec:intro}
Number theory is the branch of mathematics that studies the ring of integer numbers $\mathbb{Z}$ and its field of fractions $\mathbb{Q}$, the rational numbers. While this description may seem deceptively simple, it is a very rich area, involving myriads of abstractions and techniques. 

Consider for example the problem of finding all integer solutions to a polynomial equation in several variables (a `Diophantine equation'). Perhaps the most famous of these equations is $x^n + y^n = z^n$, where $n$ is an integer greater than $2$. Fermat's Last Theorem tells us that this equation has no integer solutions for which the product $xyz$ is nonzero. While Fermat was able to state this conjecture around 1637, its proof was not concluded until 1995, although some particular cases were established sooner.

The general proof, due to Wiles and Taylor, is built upon the combined work of hundreds of mathematicians who over the last couple of centuries developed a rich arithmetic theory of elliptic curves, modular forms and Galois representations. The key result is a special case of the Taniyama--Shimura--Weil conjecture. If we want to be able to formalize a complete proof of Fermat's Last Theorem in a theorem prover, we first need to formalize all the necessary ingredients.

In this paper we formalize the ring of ad\`eles and the group of id\`eles of a \textit{global field} (a generalization of the field $\mathbb{Q}$). As a consequence of our work we are able to state the main theorem of global class field theory. Class field theory is needed for the proof of the Taniyama--Shimura--Weil conjecture, which implies Fermat's Last Theorem. Ad\`eles and id\`eles are used in many areas of current research, including the theory of automorphic forms and the Langlands program, an ambitious group of conjectures that seek to establish deep connections between geometry and number theory.

\iffalse
\textcolor{red}{The general proof, due to Wiles and Taylor, is built upon the combined work of hundreds of mathematicians who over the last couple of centuries developed a rich arithmetic theory of elliptic curves, modular forms and Galois representations. The key result is the Taniyama--Shimura--Weil conjecture, also known as the Modularity Theorem, whose proof requires class field theory. If we want to be able to formalize a complete proof of Fermat's Last Theorem in a theorem prover, we first need to formalize all the necessary ingredients.}

\textcolor{red}{In this paper we formalize the ring of ad\`eles and the group of id\`eles of a \textit{global field} (a generalization of the field $\mathbb{Q}$). As a consequence of our work we are able to state the main theorem of global class field theory, which is a key input into the proof of Fermat's last theorem. Ad\`eles and id\`eles are used in many areas of current research, including the Langlands program, an ambitious group of conjectures that seek to establish deep connections between geometry and number theory.} %Both the Taniyama-Shimura-Weil conjecture and class field theory are part of the Langlands program.}
\fi

Our formalization was carried out using the Lean 3 theorem prover \cite{Lean3}. At the time of writing this paper, the source code is in the process of being integrated in Lean's mathematics library \texttt{mathlib}. We provide a public repository\footnote{\url{https://github.com/mariainesdff/ideles/tree/journal-submission}} containing the version of the code referred to in this article and the associated documentation\footnote{\url{https://mariainesdff.github.io/ideles/journal-submission/}} in HTML format.
Note that this is the first time that ad\`eles and id\`eles have been formalized in any theorem prover.

Before describing our formalization, we give a quick overview of the ring of ad\`eles of $\mathbb{Q}$. When studying the rational numbers, both algebraic and analytic methods can be employed. A natural way to do analysis over $\mathbb{Q}$ is by regarding it as a subspace of the real numbers $\mathbb{R}$, which are by definition the completion of $\mathbb{Q}$ with respect to the usual absolute value. However, this is not the only absolute value that can be defined on $\mathbb{Q}$: in fact, for every prime number $p$, there is a $p$-adic absolute value $|\cdot|_p$ and we can consider the corresponding completion $\mathbb{Q}_p$ of $\mathbb{Q}$. Ostrowki's theorem tells us that, up to equivalence, there are no more nontrivial absolute values on the rational numbers.

We remark that while the field $\mathbb{Q}_p$ of $p$-adic numbers is a basic object in number theory, it was not formalized in any proof assistant until 2015, when Pelayo, Voevodsky, and Warren formalized it in the Coq UniMath library \cite{padicsCoq}. The $p$-adic numbers were added to Lean's mathematical library \texttt{mathlib} in 2018, by R. Y. Lewis \cite{padicsLean}.

Since the various absolute values on $\mathbb{Q}$ provide us with different insights about the rationals, a natural question is whether it is possible to study all of them simultaneously. A first approximation would be to consider the product of the completions with respect to each absolute value. However, for technical reasons it is better to work with the following subset of the product:
\[ \mathbb{A}_\mathbb{Q} := \sideset{}{'}\prod_{p} \mathbb{Q}_p \times \mathbb{R} := \left\{ ((x_p)_p, r) \in \sideset{}{}\prod_{p} \mathbb{Q}_p \times \mathbb{R} \:\middle|\: |x_p|_p \le 1 \text{ for all but finitely many } p \right\}. \]

$\mathbb{A}_\mathbb{Q}$ is a ring under component-wise addition and multiplication, it contains $\mathbb{Q}$ as a subring via the diagonal map $r \mapsto ((r)_p, r)$, and it can be endowed with a topology that makes it into a locally compact topological ring. We call $\mathbb{A}_\mathbb{Q}$ the ring of ad\`eles or ad\`ele ring of $\mathbb{Q}$ and $\mathbb{A}_{\mathbb{Q}, f} := {\sideset{}{'_p}\prod} \mathbb{Q}_p$ its finite ad\`ele ring. The groups of units of these rings are respectively called the id\`ele group $\mathbb{I}_\mathbb{Q}$ and finite id\`ele group $\mathbb{I}_{\mathbb{Q}, f}$ of $\mathbb{Q}$.

The definitions of ad\`ele ring and id\`ele group can be generalized to any global field $K$ \cite{Global}; see sections \ref{sec:ad-id} and \ref{sec:ad-id-ff} for the details. Global fields are one of the main subjects of study in algebraic number theory and they can be of two kinds: number fields, which are finite extensions of the field $\mathbb{Q}$, and function fields, which are finite extensions of the field $\mathbb{F}_q(t)$ of rational functions over a finite field $\mathbb{F}_q$. 

Every global field is the field of fractions of a Dedekind domain, but the converse is not true. However, the definition of finite ad\`ele ring makes sense for any Dedekind domain, so we have formalized it in that degree of generality.

\subsection{Lean and mathlib}
\label{sec:Lean}
Lean 3 is a functional programming language and interactive theorem prover \cite{Lean3} based on dependent type theory, with proof irrelevance and non-cumulative universes \cite{TypeTheory}. For an introduction to Lean, see for instance \cite{TPIL}.

This project is based on Lean's mathematical library \texttt{mathlib}, which is characterized by its decentralized nature with over 300 contributors. Due to the distributed organization of \texttt{mathlib}, it is impossible to cite every author who contributed a piece of code that we used. However, we remark that our formalization makes extensive use of the theory of Dedekind domains \cite{DedekindLean} and of the theory of uniform spaces and completions, originally developed in the perfectoid space formalization project \cite{PerfectoidLean}.

In Lean's core library and \texttt{mathlib}, type classes are used to handle mathematical structures on types. For example, the type class \texttt{ring} packages two operations, addition and multiplication, as well as a list of properties they must satisfy. Then, given a type \texttt{R}, we can declare an instance \texttt{[ring R]}, and Lean's instance resolution procedure will infer that \texttt{R} has a ring structure.
Besides \texttt{instance}, whose behaviour we have just described, we use in this paper the keywords \texttt{variables}, \texttt{def}, \texttt{lemma} and \texttt{theorem}, which have the evident meaning.

\subsection{Structure of the paper}
\label{sec:cont}
We start Section \ref{sec:fin} with some background on Dedekind domains and their nonarchimedean absolute values, which we then use to define the finite ad\`ele ring and the finite id\`ele group and explore how the latter is related to the group of invertible fractional ideals. In Section \ref{sec:ad-id}, we build on this work to define the ad\`ele ring, the id\`ele group and the id\`ele class group of a number field, while in Section \ref{sec:ad-id-ff} we treat the function field case. In Section \ref{sec:cft} we discuss two applications of the id\`ele group to class field theory. Finally, we conclude Section \ref{sec:disc} with some implementation remarks and a discussion of future work connected to this project.
\section{The finite ad\`ele ring of a Dedekind domain}
\label{sec:fin}
\subsection{Dedekind domains and adic valuations}
\label{sec:Ded}
There are several equivalent definitions of Dedekind domain, three of which have been formalized in \texttt{mathlib} \cite{DedekindLean}. We work with the one formalized in \texttt{is\_dedekind\_domain} : a Dedekind domain $R$ is an integrally closed Noetherian integral domain with Krull dimension 0 or 1 \cite{Neukirch99}.

A Dedekind domain of Krull dimension 0 is a field. In this project we will only consider Dedekind domains of Krull dimension 1, for which the maximal ideals are exactly the nonzero prime ideals. Some examples are the integers $\mathbb{Z}$, the Gaussian integers $\mathbb{Z}[i] := \{ a + bi \mid a, b \in \mathbb{Z}\}$, or the ring of univariate polynomials $k[X]$ over a field $k$. All of these examples are unique factorization domains; however, not every Dedekind domain is. For instance, $\mathbb{Z}[\sqrt{-5}] := \{ a + b\sqrt{-5} \mid a, b \in \mathbb{Z}\}$ is a Dedekind domain but not a unique factorization domain, since elements like $6 = 2\cdot3 = (1 + \sqrt{-5})\cdot(1 - \sqrt{-5})$ admit two genuinely distinct factorizations.

The maximal spectrum of $R$ is the set of its maximal ideals (implemented as a type in Lean). The fraction field $K$ of $R$ is the smallest field containing $R$; its elements can be represented by fractions $r/s$, where $r$ and $s$ are in $R$ and $s$ is nonzero. For example, the fraction fields of 
$\mathbb{Z}$, $\mathbb{Z}[i]$, and $k[X]$ are respectively $\mathbb{Q}$, $\mathbb{Q}(i):= \{ a + bi \mid a, b \in \mathbb{Q}\}$, and the field $k(X)$
of rational functions over $k$.

\begin{lstlisting}[mathescape]
variables (R : Type*) [comm_ring R] [is_domain R] [is_dedekind_domain R]
	{K : Type*} [field K] [algebra R K] [is_fraction_ring R K] 
-- Note : not the maximal spectrum if R is a field
def maximal_spectrum := {v : prime_spectrum R // v.val ≠ 0 }
variable (v : maximal_spectrum R)
\end{lstlisting}

Let $R$ be a Dedekind domain (of Krull dimension 1). Then every nonzero ideal of $R$ can be written as a product of maximal ideals, and this factorization is unique up to reordering. In particular, given an element $r \in R$ and a maximal ideal $v$ of $R$, we can count how many times $v$ appears in the factorization of the principal ideal $(r)$, and this defines a nonarchimedean additive valuation on $R$ \cite[Chapter II]{Janusz}, that is, a function $\text{val}_v : R \to \mathbb{Z} \cup \{\infty\}$ such that
\begin{enumerate}
	\item $\text{val}_v(r) = \infty$ if and only if $r = 0$,
	\item $\text{val}_v (rs) = \text{val}_v(r) + \text{val}_v(s)$ for all $r, s$ in $R$, and
	\item $\text{val}_v (r + s) \geq \min \{\text{val}_v(r), \text{val}_v(s)\}$ for all $r, s$ in $R$.
\end{enumerate}
The function $\text{val}_v$ is called the $v$-adic valuation on $R$. It can be extended to a valuation on the fraction field $K$ of $R$ by defining $\text{val}_v(r/s) := \text{val}_v(r) - \text{val}_v(s)$. For example, when $R = \mathbb{Z}$ and $v = (p)$ is the ideal generated by a prime number, $\text{val}_v$ is the $p$-adic valuation on $\mathbb{Z}$ and $\mathbb{Q}$.

For both theoretical and implementation reasons, it is more convenient to work with the multiplicative version of the valuation: given any real number $n_v > 1$, we define a function
$ |\cdot|_v: R \to n_v^{\mathbb{Z} \cup \{-\infty\}} = n_v^\mathbb{Z} \cup \{0\}$ sending $r$ to $n_v^{-\text{val}_v(r)}$. From the definition of $\text{val}_v$, we immediately deduce that $|\cdot|_v$ has the following properties:
\begin{romanenumerate}
	\item $|r|_v = 0$ if and only if $r = 0$,
	\item $|rs|_v = |r|_v|s|_v$ for all $r, s$ in $R$, and
	\item $|r + s|_v \leq \max \{|r|_v, |s|_v\}$ for all $r, s$ in $R$.
\end{romanenumerate}
A function $|\cdot|_v$ satisfying conditions (i) -- (iii) is called a nonarchimedean absolute value (note that the third condition is stronger than $|r + s|_v \leq |r|_v + |s|_v$). The choice of $n_v$ used in the definition is not relevant, in the sense that any two choices of $n_v$ will yield equivalent absolute values. If, instead of property (iii), the function $|\cdot|_v$ satisfies the weaker condition $|r + s|_v \leq |r|_v + |s|_v$, we say that it is an archimedean absolute value.

We formalized the $v$-adic absolute value on $R$ in \texttt{mathlib} using the structure \texttt{valuation}, which consists on a function $|\cdot|$ from a ring $R$ to a \texttt{linear\_ordered\_comm\_monoid\_with\_zero} $\Gamma_0$ satisfying conditions (ii) and (iii), plus $|0| = 0$ and $|1| = 1$.
We chose $\Gamma_0$ equal to \texttt{with\_zero (multiplicative $\mathbb{Z}$)}, which is a way to represent $n_v^\mathbb{Z} \cup \{0\}$ in Lean. We used \texttt{associates.mk} instead of working directly with ideals simply because the corresponding factorization API was more convenient.
\begin{lstlisting}[mathescape,label=list:8-6,captionpos=t,abovecaptionskip=-\medskipamount]
def int_valuation_def (r : R) : with_zero (multiplicative $\mathbb{Z}$) :=
ite (r = 0) 0 (multiplicative.of_add (-(associates.mk v.val.val).count
  (associates.mk (ideal.span{r} : ideal R)).factors : $\mathbb{Z}$))
def int_valuation (v : maximal_spectrum R) : 
	valuation R (with_zero (multiplicative ℤ)) :=
{ to_fun    := v.int_valuation_def, 
  map_zero' := int_valuation.map_zero' v,
  map_one'  := int_valuation.map_one' v,
  map_mul'  := int_valuation.map_mul' v,
  map_add'  := int_valuation.map_add' v }
\end{lstlisting}

We extended \texttt{int\_valuation} to a \texttt{valuation} on the fraction field $K$, by setting the valuation of a fraction to be the valuation of the numerator divided by the valuation of the denominator. We checked in \texttt{lemma valuation\_well\_defined} that this definition does not depend on the choice of fraction used to represent an element of $K$.

\begin{lstlisting}[mathescape]
def valuation_def (x : K) : (with_zero (multiplicative ℤ)) :=
let s := classical.some (classical.some_spec (is_localization.mk'_surjective (non_zero_divisors R) x)) in
	 (v.int_valuation_def (classical.some (is_localization.mk'_surjective (non_zero_divisors R) x)))/(v.int_valuation_def s)
\end{lstlisting}

\begin{lstlisting}[mathescape]
lemma valuation_well_defined {r r' : R} {s s' : non_zero_divisors R} 
	(h_mk : is_localization.mk' K r s = is_localization.mk' K r' s') :
	(v.int_valuation_def r)/(v.int_valuation_def s) =
	(v.int_valuation_def r')/(v.int_valuation_def s')  
\end{lstlisting}

\iffalse
\begin{lstlisting}[mathescape]
lemma valuation_of_mk' {r : R} {s : non_zero_divisors R} :
	v.valuation_def (is_localization.mk' K r s) =
	(v.int_valuation_def r)/(v.int_valuation_def s) :=
begin
	rw valuation_def,
	exact valuation_well_defined K v
 	  (classical.some_spec (classical.some_spec (is_localization.mk'_surjective
 	    (non_zero_divisors R) (is_localization.mk' K r s)))),
end
\end{lstlisting}
\begin{lstlisting}[mathescape]
lemma valuation_of_algebra_map {r : R} :
v.valuation_def (algebra_map R K r) = v.int_valuation_def r :=
by rw [← is_localization.mk'_one K r, valuation_of_mk', submonoid.coe_one, int_valuation.map_one', div_one _]
lemma valuation_le_one (r : R) : v.valuation_def (algebra_map R K r) ≤ 1 :=
by {rw valuation_of_algebra_map, exact v.int_valuation_le_one r}
lemma valuation_lt_one_iff_dvd (r : R) : v.valuation_def (algebra_map R K r) < 1 ↔ v.val.val | ideal.span {r} :=
by { rw valuation_of_algebra_map,
exact v.int_valuation_lt_one_iff_dvd r }
\end{lstlisting}
\fi
We proved several properties of the valuation, of which we remark the fact that for every maximal ideal $v$ of $R$, there exists a uniformizer $\pi_v \in K$ for the $v$-adic valuation, that is, an element having absolute value $ |\pi_v|_v = n_v^{-1}$, or equivalently additive $v$-adic valuation $1$. 
\begin{lstlisting}[mathescape]
lemma valuation_exists_uniformizer :
	∃ (π : K), v.valuation_def π = multiplicative.of_add (-1 : ℤ)
\end{lstlisting}

Since $|\cdot|_v$ is an absolute value on the Dedekind domain $R$ and its field of fractions $K$, we can complete $R$ and $K$ with respect to $|\cdot|_v$. We denote the respective completions by $R_v$ and $K_v$, and recall that $R_v$ is an integral domain with field of fractions $K_v$.

We first formalize the definition of $K_v$ using the theory of completions of valued fields available in \texttt{mathlib}, which was originally developed as part of the formalization of perfectoid spaces \cite{PerfectoidLean}. Among the possible ways to define $K_v$, this one was chosen because of its powerful API : we can use the \texttt{field\_completion} instance to recover the fact that $K_v$ is a field, and \texttt{valued.extension\_valuation} to extend the $v$-adic valuation on $K$ to a valuation on the completion $K_v$.

\begin{lstlisting}[mathescape]
def v_valued_K (v : maximal_spectrum R) : valued K := 
{ Γ₀  := (with_zero (multiplicative ℤ)),
  grp := infer_instance,
  v   := v.valuation }
\end{lstlisting}

\begin{lstlisting}[mathescape]
def K_v := @uniform_space.completion K (us' v)
instance : field (K_v K v) := @field_completion K _ (us' v) (tdr' v) _ (ug' v)
instance valued_K_v : valued (K_v K v) := 
{ Γ₀  := with_zero (multiplicative ℤ),
  grp := infer_instance,
  v   := @valued.extension_valuation K _ (v_valued_K v) }
\end{lstlisting}

It can be shown that $R_v$ is equal to the ring of integers of $K_v$, that is, the subring of $K_v$ consisting of elements of absolute value less than or equal to one. In our formalization, we actually use this characterization to define $R_v$, so that we automatically have an inclusion of $R_v$ in $K_v$.

\begin{lstlisting}[mathescape]
def R_v : subring (K_v K v) := 
@valuation.integer (K_v K v) (with_zero (multiplicative ℤ)) _ _ (valued_K_v v).v 
\end{lstlisting}

\subsection{The finite ad\`ele ring}
\label{sec:fin-ad}
Now that we have defined nonarchimedean absolute values on a Dedekind domain $R$ and their extension to $K$, we can attempt to 
simultaneously study all of them. In order to do so, we define the finite ad\`ele ring $\mathbb{A}_{R, f}$ of $R$ as the restricted product of the completions $K_v$ with respect to their ring of integers $R_v$, i. e.,
\[ \mathbb{A}_{R, f} := {\prod_{v}}^{'} K_v := \left\{(x_v)_v \in \prod_{v} K_v \:\middle|\: x_v \in R_v \text{ for all but finitely many } v  \right\},  \]
where $v$ runs over the set of maximal ideals of $R$. Recall that $x_v \in R_v$ is equivalent to $|x_v|_v \leq 1$, so $\mathbb{A}_{R, f}$ is an immediate generalization of $\mathbb{A}_{\mathbb{Q}, f}$.

Since $\mathbb{A}_{R, f}$ is a subset of the product $\prod_{v} K_v$, it is easy to prove that it is a commutative ring with component-wise addition and multiplication (one just needs to check that it is closed under addition, negation and multiplication).

\begin{lstlisting}[mathescape]
def K_hat := (Π (v : maximal_spectrum R), (K_v K v))
def finite_adele_ring' := { x : (K_hat R K) // $\forall^f$ (v : maximal_spectrum R) in filter.cofinite, (x v ∈ R_v K v) }
instance : comm_ring (finite_adele_ring' R K) := ...
\end{lstlisting}

We endow $\mathbb{A}_{R, f}$ with the topology generated by the set $\{\Pi_v U_v \:|\: U_v \text{ is open and } U_v = R_v \text{ for almost all } v \}$ and prove that addition and multiplication on $\mathbb{A}_{R, f}$ are continuous for this topology, which makes $\mathbb{A}_{R, f}$ into a topological ring. While these proofs are not conceptually hard, their formalization turned out to be quite long.

\begin{lstlisting}[mathescape]
def finite_adele_ring'.generating_set : set (set (finite_adele_ring' R K)) := 
{U : set (finite_adele_ring' R K) |
	∃ (V : Π (v : maximal_spectrum R), set (K_v K v)),
	  (∀ x : finite_adele_ring' R K, x ∈ U ↔ ∀ v, x.val v ∈ V v) ∧
	  (∀ v, is_open (V v)) ∧ $\forall^f$ v in filter.cofinite, V v = R_v K v} 
instance : topological_space (finite_adele_ring' R K) :=
topological_space.generate_from (finite_adele_ring'.generating_set R K)
\end{lstlisting}

\iffalse
\begin{lstlisting}[mathescape]
lemma finite_adele_ring'.is_open_integer_subring :
is_open {x : finite_adele_ring' R K | ∀ (v : maximal_spectrum R), x.val v ∈ R_v K v} :=
begin  
apply topological_space.generate_open.basic,
rw finite_adele_ring'.generating_set,
use λ v, R_v K v,
refine ⟨λ v, by refl, λ v, K_v.is_open_R_v R K v,_⟩,
{ rw filter.eventually_cofinite,
simp_rw [eq_self_iff_true, not_true, set_of_false, finite_empty] },
end
\end{lstlisting}
\fi

For every element $k \in K$, there are finitely many maximal ideals $v$ of $R$ such that the $v$-adic absolute value of $k$ is greater than 1; hence $(k)_v$ is a finite ad\`ele of $R$. The map $\text{inj}_K : K \to \mathbb{A}_{R, f}$ sending $k$ to $(k)_v$ is an injective ring homomorphism, which allows us to regard $K$ as a subring of $\mathbb{A}_{R, f}$.

\begin{lstlisting}[mathescape]
def inj_K : K → finite_adele_ring' R K := 
λ x, ⟨(λ v : maximal_spectrum R, (coe : K → (K_v K v)) x), inj_K_image R K x⟩
\end{lstlisting}

One might wonder why we defined $\mathbb{A}_{R, f}$, instead of just working with the full product $\prod_{v} K_v$. The main reason for this is that, while both $\mathbb{A}_{R, f}$ and $\prod_{v} K_v$ are topological rings containing $K$ as a subring, only the former is locally compact and contains $K$ as a discrete and co-compact subring. Since $\mathbb{A}_{R, f}$ is in particular a locally compact topological group, it is possible to define a (unique up to scalars) Haar measure on $\mathbb{A}_{R, f}$, which allows us to integrate functions over $\mathbb{A}_{R, f}$. Tate famously used this integration theory in his thesis to study the properties of Hecke $L$-functions of number fields. Note that Haar measures have recently been formalized in \texttt{mathlib} \cite{HaarLean}.

\subsubsection{Alternative definition of the finite ad\`ele ring}
\label{sec:alt-def}
There is a second characterization of the ring of finite ad\`eles of $R$ which is also widely used in number theory. We start with the product $\hat{R} := \prod_{v} R_v$ over all maximal ideals of $R$ and observe that it contains $R$ via the diagonal inclusion $r \mapsto (r)_v$. Hence, we can consider the localization $(\prod_{v} R_v)[\frac{1}{R\setminus{\{0\}}}]$ of $\hat{R}$ at $R\setminus{\{0\}}$, consisting of tuples of the form $(\frac{r_v}{s})_v$ where $r_v \in R_v$ for all $v$ and $s \in R\setminus{\{0\}} \subseteq R_v\setminus{\{0\}}$.

To define the topological ring structure on $\hat{R}[\frac{1}{R\setminus{\{0\}}}]$, we use the fact that for any ring $S$, ring topologies on $S$ form a complete lattice. In particular, given any map $f : T \to S$ from a topological space $T$ to a ring $S$, one can define the coinduced ring topology on $S$ to be the finest topology such that $S$ is a topological ring and $f$ is continuous. The complete lattice structure was formalized as part of this project and is already a part of \texttt{mathlib}.
We give $\hat{R}[\frac{1}{R\setminus{\{0\}}}]$ the ring topology coinduced by the localization map $(r_v)_v \mapsto (\frac{r_v}{1})_v$ from
$\hat{R}$ with the product topology to $\hat{R}[\frac{1}{R\setminus{\{0\}}}]$.

It is well known that $\mathbb{A}_{R, f}$ is isomorphic to $(\prod_{v} R_v)[\frac{1}{R\setminus{\{0\}}}]$ as topological rings. Given an element $(\frac{r_v}{s})_v \in (\prod_{v} R_v)[\frac{1}{R\setminus{\{0\}}}]$, the absolute value $|\frac{r_v}{s}|_v$ will be less than or equal to one, except possibly at the finitely many $v$ dividing the denominator $s$; hence $(\frac{r_v}{s})_v$ is a finite ad\`ele and one easily sees that this map is an isomorphism of rings. Checking that it is also a homeomorphism requires more work.

We formalized this second definition of the ad\`ele ring in \texttt{finite\_adele\_ring}, but we omit for now the formalization of the proof that the two definitions yield isomorphic topological rings. The \texttt{finite\_adele\_ring} definition has the advantage that, being defined as a localization, \texttt{finite\_adele\_ring R} automatically inherits a commutative topological ring structure, while for \texttt{finite\_adele\_ring' R} this has to be proven by hand. However, we found that for proving results such as the one described in Section \ref{sec:surj}, our first definition was easier to work with.

\begin{lstlisting}[mathescape]
def finite_adele_ring := localization (diag_R R K)
instance : comm_ring (finite_adele_ring R K) := localization.comm_ring
instance : algebra (R_hat R K) (finite_adele_ring R K) := localization.algebra
instance : is_localization (diag_R R K) (finite_adele_ring R K):=
localization.is_localization
instance : topological_space (finite_adele_ring R K) :=
localization.topological_space
instance : topological_ring (finite_adele_ring R K) :=
localization.topological_ring
\end{lstlisting}

\subsection{The finite id\`ele group}
\label{sec:fin-id}
The finite id\`ele group $\mathbb{I}_{R, f}$ of $R$ is the unit group of the finite ad\`ele ring $\mathbb{A}_{R, f}$.
It is a topological group with the topology induced by the map $\mathbb{I}_{R, f} \to \mathbb{A}_{R, f} \times \mathbb{A}_{R, f}$ sending $x$ to $(x, x^{-1})$.

\begin{lstlisting}[mathescape]
def finite_idele_group' := units (finite_adele_ring' R K)
instance : topological_space (finite_idele_group' R K) := units.topological_space
instance : group (finite_idele_group' R K) := units.group
instance : topological_group (finite_idele_group' R K) := units.topological_group
\end{lstlisting}

Note that for every nonzero $k \in K$, the finite ad\`ele $(k)_v$ is invertible, with inverse $(k^{-1})_v$. It follows that $\mathbb{I}_{R, f}$ contains $K^* = K\setminus \{0\}$ as a subgroup. We formalize this fact by defining a function \texttt{inj\_units\_K} from $K^*$ to $\mathbb{I}_{R, f}$ and proving that it is an injective group homomorphism.
\begin{lstlisting}[mathescape]
def inj_units_K : units K → finite_idele_group' R K := 
λ x, ⟨inj_K R K x.val, inj_K R K x.inv, right_inv R K x, left_inv R K x⟩
\end{lstlisting}

\subsection{Relation to fractional ideals}
\label{sec:frac}

The finite id\`ele group of $R$ is closely related to its group of invertible fractional ideals. A fractional ideal of $R$ is an $R$-submodule $I$ of $K$ for which there exists an $a \in R$ such that $aI$ is an ideal $J$ of $R$. We say that $I$ is invertible if there exists another fractional ideal $I'$ such that $II' = R$. 

For a Dedekind domain $R$, every nonzero fractional ideal is invertible and can be factored as a product $v_1^{n_1} \cdots v_m^{n_m}$ of maximal ideals of $R$ where the $n_i$ are integers, uniquely up to reordering of the factors. We formalized this definition in \texttt{fractional\_ideal.factorization}, where we express $I$ as a \texttt{finprod} over all maximal ideals of $R$. We also provide some API to work with the exponents appearing in this factorization.
\begin{lstlisting}[mathescape]
lemma fractional_ideal.factorization (I : fractional_ideal (non_zero_divisors R) K)
  (hI : I ≠ 0) {a : R} {J : ideal R}
  (haJ : I = fractional_ideal.span_singleton (non_zero_divisors R) 
    ((algebra_map R K) a)⁻¹ * ↑J) :
  $\Pi^f$ (v : maximal_spectrum R),
    (v.val.val : fractional_ideal (non_zero_divisors R) K)^ ((associates.mk v.val.val).count (associates.mk J).factors - (associates.mk v.val.val).count (associates.mk (ideal.span{a})).factors : ℤ) = I
\end{lstlisting}

We can define a group homomorphism from $\mathbb{I}_{R,f}$ to the group of invertible fractional ideals by sending $ (x_v)_v \in \mathbb{I}_{R, f}$ to the product $\Pi_v v^{\text{val}_v(x_v)}$. Since for every $(x_v)_v \in \mathbb{I}_{R, f}$ there are finitely many maximal ideals $v$ such that $\text{val}_v(x_v)$ is nonzero, this product is actually finite, so it indeed defines a nonzero fractional ideal of $R$.
\begin{lstlisting}[mathescape]
def finite_idele.to_add_valuations (x : finite_idele_group' R K) :
	Π (v : maximal_spectrum R), ℤ := λ v, -(with_zero.to_integer ((valuation.ne_zero_iff valued.v).mpr (v_comp.ne_zero R K v x)))
lemma finite_add_support (x : finite_idele_group' R K ) :  
	$\forall^f$ (v : maximal_spectrum R) in filter.cofinite, 
	  finite_idele.to_add_valuations R K x v = 0 := ...
def map_to_fractional_ideals.val :
	(finite_idele_group' R K) → (fractional_ideal (non_zero_divisors R) K) := 
λ x, $\Pi^f$ (v : maximal_spectrum R), (v.val.val : fractional_ideal (non_zero_divisors R) K)^(finite_idele.to_add_valuations R K x v)
\end{lstlisting}

We show that this homomorphism is surjective and its kernel is the set $\mathbb{I}_{R, \infty}$ of elements $(x_v)_v$ in $\mathbb{I}_{R,f}$ having additive valuation zero at all $v$. Moreover, this map is continuous when the group of invertible fractional ideals is given the discrete topology.

\section{Ad\`eles and id\`eles of number fields}
\label{sec:ad-id}
\subsection{Number fields and their rings of integers}
\label{sec:nf}
A number field $K$ is a finite extension of the field $\mathbb{Q}$ of rational numbers \cite{Janusz}. Every finite extension is algebraic, so every element $k \in K$ is the root of a polynomial with coefficients in $\mathbb{Q}$. If moreover $k$ is the root of a monic polynomial with integer coefficients, we say that $k$ is an algebraic integer. The algebraic integers of $K$ form a subring $\mathcal{O}_K$, called the ring of integers of $K$, which is a Dedekind domain of Krull dimension 1 in which every nonzero ideal is of finite index.

Remember from the introduction that one motivation for defining the ad\`eles of $K$ was to simultaneously study all the (equivalence classes of) nontrivial absolute values on $K$. These absolute values can be split into two kinds: nonarchimedean and archimedean. The nonarchimedean ones are exactly the $v$-adic absolute values associated to maximal ideals of the ring of integers $\mathcal{O}_K$, discussed in section \ref{sec:Ded}.

To obtain the archimedean absolute values, we first recall that we can find a $\mathbb{Q}$-vector space basis of $K$ of the form $\{ 1, \alpha, \dots, \alpha^{n-1}\}$, where $n$ is the dimension of $K$ over $\mathbb{Q}$ and $\alpha$ is an element of $K$. This $\alpha$ is a root of a degree $n$ polynomial $f_\alpha$ with coefficients in $\mathbb{Q}$. For each real root $r$ of $f_\alpha$, we get an embedding of $K$ into the real numbers $\mathbb{R}$ (the map sending $\alpha$ to $r$), and restricting the usual absolute value on $\mathbb{R}$ to the image of $K$, we get an archimedean absolute value on $K$. Similarly, for every pair of complex conjugate roots $(s_1, s_2)$ of $f_\alpha$, we get a pair of embeddings of $K$ into the complex numbers $\mathbb{C}$, and we can restrict the complex absolute value to the image of $K$ under one of them to get an absolute value on $K$. Note that the two embeddings coming from a conjugate pair yield equivalent absolute values.

\subsection{The ring of ad\`eles}
\label{sec:ad}
Let $K$ be a number field. We define the ring of ad\`eles of $K$ as the restricted product of the completions $K_v$ of $K$ with respect to each absolute value $|\cdot|_v$ on it:
$ \mathbb{A}_K := \sideset{}{'_{|\cdot|_v}}\prod K_v$. That is, $\mathbb{A}_K$ is the subring of the product $\sideset{}{_{|\cdot|_v}}\prod K_v$ consisting on tuples $(a_v)_v$ such that $|a_v|_v \leq 1$ for all but finitely many $v$.
Since each nonarchimedean absolute value $|\cdot|_v$ corresponds to a maximal ideal $v$ of $O_K$, and there are finitely many archimedean absolute values, we can rewrite this definition as 
\[ \mathbb{A}_K := \sideset{}{'}\prod_{v \text{ max.}} {K_v} \times \sideset{}{}\prod_{|\cdot|_v \text{ arch.}} {K_v} = \sideset{}{'}\prod_{v \text{ max.}} {K_v}\times (\mathbb{R} \otimes_\mathbb{Q} K),  \]
where we have used a theorem from algebraic number theory to get the second equality. Note that $\sideset{}{'_{v}}\prod K_v$ is the finite ad\`ele ring associated to the Dedekind domain $\mathcal{O}_K$; we will denote it by $\mathbb{A}_{K, f}$ and call it the finite ad\`ele ring of $K$. We formalize these definitions as follows:
\begin{lstlisting}[mathescape]
variables (K : Type) [field K] [number_field K]
def A_K_f := finite_adele_ring' (ring_of_integers K) K
def A_K := (A_K_f K) × (ℝ ⊗[ℚ] K)
\end{lstlisting}

We proved in Section \ref{sec:fin-ad} that \texttt{A\_K\_f} is a topological commutative ring. The product and tensor product of commutative rings are commutative rings, so \texttt{A\_K} is a commutative ring. To prove that it is a topological commutative ring, it therefore suffices to show that $\mathbb{R} \otimes_\mathbb{Q} K$ is a topological ring.
We do this by using the fact that there are isomorphisms $\mathbb{R}^n \simeq \mathbb{R} \otimes_\mathbb{Q} \mathbb{Q}^n \simeq \mathbb{R} \otimes_\mathbb{Q} K$, where $n$ is the dimension of $K$ over $\mathbb{Q}$.

Note that $\mathbb{R}^n$ is represented in Lean by the type \texttt{fin n → $\mathbb{R}$} of functions from $\{1, \dots, n\}$ to $\mathbb{R}$, and we can use \texttt{pi} to get its topological commutative ring structure as follows:

\begin{lstlisting}[mathescape]
variables (n : ℕ) 
instance : ring (fin n → ℝ) := pi.ring
instance : topological_space (fin n → ℝ) := Pi.topological_space
instance : has_continuous_add (fin n → ℝ) := pi.has_continuous_add'
instance : has_continuous_mul (fin n → ℝ) := pi.has_continuous_mul'
instance : topological_ring (fin n → ℝ) := topological_ring.mk
\end{lstlisting}

 We then define the topology on $\mathbb{R} \otimes_\mathbb{Q} K$ as the ring topology coinduced by the map $ \mathbb{R}^n \to \mathbb{R} \otimes_\mathbb{Q} K$, where $\mathbb{R}^n$ has the product topology.
 Finally, \texttt{A\_K} becomes a topological ring with the product topology.

\begin{lstlisting}[mathescape]
def linear_map.Rn_to_R_tensor_K : (fin (finite_dimensional.finrank ℚ K) → ℝ) →$_{\texttt{l}}$[ℝ] (ℝ ⊗[ℚ] K) := 
linear_map.comp (linear_map.base_change K) (linear_map.Rn_to_R_tensor_Qn K)
def infinite_adeles.ring_topology : ring_topology (ℝ ⊗[ℚ] K) := 
ring_topology.coinduced (linear_map.Rn_to_R_tensor_K K)
instance : topological_space (ℝ ⊗[ℚ] K) :=
(infinite_adeles.ring_topology K).to_topological_space
instance : topological_ring (ℝ ⊗[ℚ] K) :=
(infinite_adeles.ring_topology K).to_topological_ring
instance : topological_space (A_K K) := prod.topological_space
instance : topological_ring (A_K K) := prod.topological_ring
\end{lstlisting}

We end this section by recalling that $\mathbb{A}_{K, f}$ contains the field $K$ as a subring, via the diagonal map sending $k \in K$ to the finite ad\`ele $(k)_v$. Combining this with the natural inclusion $ k \mapsto 1 \otimes k$ of $K$ in $\mathbb{R} \otimes_\mathbb{Q} K$, we can also view $K$ as a subring of $\mathbb{A}_{K}$.

\begin{lstlisting}[mathescape]
def inj_K_f : K → A_K_f K := inj_K (ring_of_integers K) K
def inj_K : K → A_K K := 
λ x, ⟨inj_K_f K x, algebra.tensor_product.include_right x⟩
\end{lstlisting}

\subsection{The group of id\`eles and the id\`ele class group}
\label{sec:id}
We define the group $\mathbb{I}_K$ of id\`eles of $K$ as the unit group of the ring of ad\`eles $\mathbb{A}_K$, and the group $\mathbb{I}_{K, f}$ of finite id\`eles as the unit group of $\mathbb{A}_{K, f}$.
\begin{lstlisting}[mathescape]
def I_K_f := units (A_K_f K)
def I_K := units (A_K K)
\end{lstlisting}
For every nonzero $k \in K$, the finite ad\`ele $(k)_v$ is a unit (with inverse $(k^{-1})_v$), and so is the ad\`ele $((k)_v, 1 \otimes k)$. Therefore, we can regard $K^*$ as a subgroup of the (finite) id\`ele group, which allows us to define the id\`ele class group $C_K$ of $K$ as the quotient of $\mathbb{I}_K$ by $K^*$:

\begin{lstlisting}[mathescape]
def C_K := (I_K K) / (inj_units_K.group_hom K).range
\end{lstlisting}
The name id\`ele class group is justified by the close relation between $C_K$ and the ideal class group of $K$, which we discuss in section \ref{sec:surj}.

\section{Ad\`eles and id\`eles of function fields}
\label{sec:ad-id-ff}
Let $k$ be a field, $k[t]$ be the ring of polynomials in one variable over $k$ and $k(t)$ be the field of rational functions (quotients of polynomials) over $k$. A function field $F$ is a finite field extension of $k(t)$ \cite{Stich}. 

\begin{lstlisting}[mathescape]
variables (k F : Type) [field k] [field F] [algebra (polynomial k) F]
  [algebra (ratfunc k) F] [function_field k F]
  [is_scalar_tower (polynomial k) (ratfunc k) F] [is_separable (ratfunc k) F]
\end{lstlisting}

All of the absolute values that can be defined over $k(t)$ are nonarchimedean: there is one $v$-adic absolute value for each maximal ideal $v$ of $k[t]$, plus one extra absolute value, called the place at infinity $|\cdot|_\infty$, defined by setting $\left|\frac{f}{g}\right|_\infty = q^{\text{deg}(f) - \text{deg}(g)}$, where $q > 1$ is a fixed real number. \iffalse and $\frac{f}{g}$ is a quotient of polynomials of $k[t]$, with $g$ nonzero. \fi The completion of $k(t)$ with respect to this absolute value is the field $k((t^{-1}))$ of Laurent series in $t^{-1}$.

Following the strategy from Section \ref{sec:Ded}, we formalize $|\cdot|_\infty$ in Lean under the name \texttt{infty\_valuation} and we let \texttt{kt\_infty} denote the completion of $k(t)$ with respect to $|\cdot|_\infty$.
\begin{lstlisting}[mathescape]
def infty_valuation_def (r : ratfunc k) : with_zero (multiplicative ℤ) :=
ite (r = 0) 0 (multiplicative.of_add ((r.num.nat_degree : ℤ) - r.denom.nat_degree))
def kt_infty := @uniform_space.completion (ratfunc k) (usq' k)
\end{lstlisting}

More generally, all of the absolute values on a function field $F$ over $k$ are nonarchimedean. Most of them correspond to maximal ideals of  the integral closure of $k[t]$ in $F$. The finite ad\`ele ring of $F$ is the restricted product
\[\mathbb{A}_{F, f} := {\sideset{}{'}\prod_{v}} F_v := \left \{ (x_v)_v \in \sideset{}{}\prod_{v} F_v \:\middle| \: \lvert x_v \rvert_v \leq 1 \text{ for all but finitely many } v \right \}, \] where $v$ runs over these maximal ideals. However, $F$ also contains a finite collection of nonarchimedean absolute values coming from the absolute value $|\cdot|_\infty$ on $k(t)$. In order to include these absolute values as well, we define the ad\`ele ring of $F$ as the product\[\mathbb{A}_F := \mathbb{A}_{F,f} \times (k((t^{-1})) \otimes_{k(t)} F).\]

\begin{lstlisting}[mathescape]
def A_F_f := finite_adele_ring' (ring_of_integers k F) F
def A_F := (A_F_f k F) × ((kt_infty k) ⊗[ratfunc k] F)
\end{lstlisting}

The (finite) ad\`ele ring of $F$ is a topological commutative ring. We define the (finite) id\`ele group of $F$ to be its group of units, respectively denoted $\mathbb{I}_{F, f}$ and $\mathbb{I}_F$, with the topology induced by the map $x \mapsto (x, x^{-1})$ as in Section \ref{sec:fin-id}.

Note that in number theory one is usually interested in the ad\`ele ring of a function field over a finite field $k = \mathbb{F}_q$. However, $\mathbb{A}_F$ can be defined for any choice of field $k$, so we do not require $k$ to be finite in our formalization; instead, this finiteness assumption will have to be included in the lemmas that need it. 

\section{Class Field Theory}
\label{sec:cft}
Class field theory is a branch of number theory whose goal is to describe the Galois abelian extensions of a local or global field $K$, as well as their corresponding Galois groups, in terms of the arithmetic of the field $K$ \cite{AT, CF, Milne}. Recall from the introduction that a global field is either a number field or a function field over a finite field $\mathbb{F}_q$. 
A local field is the completion of a global field with respect to an absolute value. Examples of local fields include the real numbers $\mathbb{R}$, the complex numbers $\mathbb{C}$, the $p$-adic numbers $\mathbb{Q}_p$, or the field $\mathbb{F}_q((X))$ of formal Laurent series over a finite field.

In this section we discuss two class field theory results involving the definition of the id\`ele class group. The first one is a proof that the ideal class group of a number field is isomorphic to a quotient of its id\`ele class group, which we describe explicitly. The second one is a formalization of the statement of the main theorem of global class field theory.
\subsection{The ideal class group is a quotient of the id\`ele class group}
\label{sec:surj}

We have seen in Section \ref{sec:frac} that, for any Dedekind domain $R$, there is a continuous surjective group homomorphism from the finite id\`ele group $\mathbb{I}_{R, f}$ to the group Fr$(R)$ of invertible fractional ideals of $R$, sending
$(x_v)_v$ to $\prod_{v} v^{\text{val}_v(x_v)}$. 

If $K$ is a number field with ring of integers $R$, we can extend this map to a group homomorphism $\mathbb{I}_K \to \text{Fr}(R)$ by pre-composing with the natural projection $\mathbb{I}_K \to \mathbb{I}_{K, f}$, obtaining again a continuous surjection. It is easy to see that an id\`ele $((x_v)_v, r \otimes_\mathbb{Q} k)\in \mathbb{I}_K$ belongs to the kernel of this map, which we denote $\mathbb{I}_{K, \infty}$, if and only if $\text{val}_v(x_v)$ is equal to zero for every maximal ideal $v$ of $R$. We wrote this map in Lean and formalized proofs of each of the listed properties.

\begin{lstlisting}[mathescape=true]
-- For a Dedekind domain R with fraction field K :
def map_to_fractional_ideals.val :
	(finite_idele_group' R K) → (fractional_ideal (non_zero_divisors R) K) :=
λ x, $\Pi_{}^{f}$ (v : maximal_spectrum R), (v.val.val : fractional_ideal (non_zero_divisors R) K)^ (finite_idele.to_add_valuations R K x v)
\end{lstlisting}

\begin{lstlisting}[mathescape=true]
lemma I_K.map_to_fractional_ideals.surjective :
	function.surjective (I_K.map_to_fractional_ideals K) := ...
lemma I_K.map_to_fractional_ideals.continuous :
	continuous (I_K.map_to_fractional_ideals K) := ...
lemma I_K.map_to_fractional_ideals.mem_kernel_iff (x : I_K K) : 
	I_K.map_to_fractional_ideals K x = 1 ↔ ∀ v : maximal_spectrum (ring_of_integers K), finite_idele.to_add_valuations (ring_of_integers K) K (I_K.fst K x) v = 0 := ...
\end{lstlisting}

Now, we want to show that this map induces a homomorphism at the level of class groups. The ideal class group Cl$(K)$ of $K$ is defined as the quotient of the group of invertible fractional ideals of $K$ by the subgroup of principal fractional ideals. It is an important object in algebraic number theory, since it can be interpreted as a measure of how far the ring of integers of $K$ is from being a unique factorization domain. 

Note that the id\`ele $((k)_v, 1 \otimes_\mathbb{Q} k)$  corresponding to a nonzero $k \in K$ gets mapped to $\prod_v v^{\text{val}_v(k)}$, which is the principal fractional ideal generated by $k$. Hence, we get an induced map from the id\`ele class group $C_K$ to the ideal class group Cl$(K)$. Using the universal property of the quotient topology, we conclude that this map $C_K \to \text{Cl}(K)$ is a continuous surjective homomorphism, with kernel $\mathbb{I}_{K, \infty}K^*/K^*$. Hence, by the first isomorphism theorem for topological groups, Cl$(K)$ is isomorphic to the quotient of $C_K$ by $\mathbb{I}_{K, \infty}K^*/K^*$.

By proving this theorem, we show that our formalization of the ad\`eles and id\`eles of a global field can be effectively used in practice to prove graduate-level number theoretic results. While we have only formalized this proof for number fields, it can be trivially adapted to the function field case.

\subsection{The main theorem of global class field theory}
\label{sec:mt}
Let $K$ be a number field, $\overline{K}$ an algebraic closure of $K$ and $G_K := \text{Gal}_{\overline{K}/ K}$ the Galois group of the extension $\overline{K}/K$. The topological group $G_K$ is isomorphic to the inverse limit $\varprojlim_{L} \text{Gal}(L/K) $ over all finite extensions $L/K$, with the inverse limit topology. We consider the topological abelianization $G_K^{ab} := G_K/\overline{[G_K, G_K]}$ of $G_K$, defined as the quotient of $G_K$ by the topological closure of the commutator subgroup of $G_K$. The group $G_K^{ab}$ is a topological group with the quotient topology, because $\overline{[G_K, G_K]}$ is a normal subgroup of $G_K$. 

An exercise in infinite Galois theory shows that $G_K^{ab}$ is the Galois group of the maximal abelian extension $K^{ab}$ of $K$. The main theorem of global class field theory allows us to describe this Galois group in terms of the id\`ele class group of $K$ :
\begin{theorem}[Main Theorem of Global Class Field Theory]
	Let $K$ be a number field. Denote by $\pi_0(C_K)$ the quotient of $C_K$ by the connected component of the identity. There is an isomorphism of topological groups $\pi_0(C_K) \simeq G_K^{\text{ab}}$.
\end{theorem}

We formalized the statement of this theorem in two parts: we first claimed the existence of a group isomorphism \texttt{main\_theorem\_of\_global\_CFT.group\_isomorphism} between $\pi_0(C_K)$ and $G_K^{ab}$ and then in \texttt{main\_theorem\_of\_global\_CFT.homeomorph} we stated that this map is also a homeomorphism. Note that a complete pen-and-paper proof of this theorem spans hundreds of pages, so we have not attempted to formalize it. 

\begin{lstlisting}[mathescape]
variables (K : Type) [field K] [number_field K]
theorem main_theorem_of_global_CFT.group_isomorphism : (number_field.C_K K) / (subgroup.connected_component_of_one (number_field.C_K K)) ≃* (G_K_ab K) :=
sorry
theorem main_theorem_of_global_CFT.homeomorph :
homeomorph ((number_field.C_K K) / (subgroup.connected_component_of_one (number_field.C_K K))) (G_K_ab K) := 
{ continuous_to_fun  := sorry,
  continuous_inv_fun := sorry,
  ..(main_theorem_of_global_CFT.group_isomorphism K) }
\end{lstlisting}

\section{Discussion}
\label{sec:disc}
\subsection{Implementation comments}
\label{sec:impl}
In this section we discuss some technical details of our formalization. The first one has to do with the universe in which the Dedekind domain $R$ and its function field $K$ are defined. To define the $v$-adic valuations and formalize the factorization of fractional ideals, we can let $R$ and $K$ be of \texttt{Type u} for any universe \texttt{u}. However, to define the completions $K_v$ and all subsequent work, we require $R$ and $K$ to live in \texttt{Type}. This is because the structure \texttt{valued}, which we used in our definitions of the completions, requires the field $K$ and the \texttt{linear\_ordered\_comm\_monoid\_with\_zero} $\Gamma_0$ to live in the same universe, and we chose $\Gamma_0$ to be \texttt{with\_zero(multiplicative $\mathbb{Z}$)}, which has type \texttt{Type}.

Secondly, we found that some definitions were unexpectedly causing timeouts or memory errors, due to the fact that Lean was not able to decide whether they were computable or not. We would like to thank Gabriel Ebner for finding the cause of these errors and providing the \texttt{force\_noncomputable} definition to address it, as well as an associated \texttt{simp} lemma.
\begin{lstlisting}[mathescape]
noncomputable def force_noncomputable {α : Sort*} (a : α) : α :=
function.const _ a (classical.choice ⟨a⟩)
@[simp] lemma force_noncomputable_def {α} (a : α) : force_noncomputable a = a :=
rfl
\end{lstlisting}
As an example, the definition of the coercion map from $\mathbb{A}_{R, f}$ to $\prod_{v}K_v$ was causing an `excessive memory consumption' error, which was immediately solved with the application of \texttt{force\_noncomputable}.
\begin{lstlisting}[mathescape]
def coe' : (finite_adele_ring' R K) → K_hat R K :=
force_noncomputable $\$$ $\lambda$ x, x.val
\end{lstlisting}

\subsection{Future work}
\label{sec:fw}
There are several natural directions for future formalization work stemming from this project. We list some of them, starting with the most immediate goals.
\begin{itemize}
\item Show that the two definitions of the finite ad\`ele ring formalized in Section \ref{sec:fin-ad} give isomorphic topological rings. Constructing an isomorphism of rings between them will be easy, but checking that it is a homeomorphism will require some work. 
\item Define the id\`ele class group and prove the results from Section \ref{sec:surj} in the function field setting. The proofs will be nearly identical to the number field case. 
\item Formalize topological results about the ad\`ele ring and the id\`ele group, such as the proof that $\mathbb{A}_K$ is locally compact and contains $K$ as a discrete co-compact subgroup.
\item Given a finite extension $L/K$ of global fields, formalize the isomorphism $\mathbb{A}_L \simeq L \otimes \mathbb{A}_K$ and its consequences.
\item Keep stating, and eventually proving, results from class field theory.
\item Formalize Tate's thesis.
\end{itemize}

More generally, having the definitions of $\mathbb{A}_K$ and $\mathbb{I}_K$ opens the door to formalizing concepts and results used in state-of-the-art number theory, including the definition of automorphic forms \cite{Bump} and the statement of the Langlands correspondence \cite{Langlands}. Note that only some cases of the Langlands correspondence have been proven, and the Langlands program is currently one of the main research areas in number theory.

%%
%% Bibliography
%%

%% Please use bibtex, 

\bibliography{ms}

\begin{thebibliography}{10}

\bibitem{AT}
Emil Artin and John Tate.
\newblock {\em Class {F}ield {T}heory}.
\newblock W. A. Benjamin, New York, 1967.

\bibitem{Global}
Emil Artin and George Whaples.
\newblock {Axiomatic Characterization of Fields by the Product Formula for
  Valuations}.
\newblock {\em Bulletin of the American Mathematical Society}, 51(7):469 --
  492, 1945.
\newblock URL:
  \url{https://mathscinet.ams.org/mathscinet-getitem?mr=MR0013145}.

\bibitem{TPIL}
Jeremy Avigad, Leonardo de~Moura, and Soonho Kong.
\newblock {\em {Theorem Proving in Lean}}.
\newblock Carnegie Mellon University, 2021. {R}elease 3.23.0.
\newblock URL: \url{https://leanprover.github.io/theorem_proving_in_lean/}.

\bibitem{DedekindLean}
Anne Baanen, Sander~R. Dahmen, Ashvni Narayanan, and Filippo A.~E. Nuccio
  Mortarino Majno~di Capriglio.
\newblock {A Formalization of Dedekind Domains and Class Groups of Global
  Fields}.
\newblock In Liron Cohen and Cezary Kaliszyk, editors, {\em 12th International
  Conference on Interactive Theorem Proving (ITP 2021)}, volume 193 of {\em
  Leibniz International Proceedings in Informatics (LIPIcs)}, pages 5:1--5:19,
  Dagstuhl, Germany, 2021. Schloss Dagstuhl -- Leibniz-Zentrum f{\"u}r
  Informatik.
\newblock URL: \url{https://drops.dagstuhl.de/opus/volltexte/2021/13900}, \href
  {https://doi.org/10.4230/LIPIcs.ITP.2021.5}
  {\path{doi:10.4230/LIPIcs.ITP.2021.5}}.

\bibitem{Bump}
Daniel Bump.
\newblock {\em {Automorphic Forms and Representations}}.
\newblock Cambridge Studies in Advanced Mathematics. Cambridge University
  Press, Cambridge, 1997.
\newblock \href {https://doi.org/10.1017/CBO9780511609572}
  {\path{doi:10.1017/CBO9780511609572}}.

\bibitem{PerfectoidLean}
Kevin Buzzard, Johan Commelin, and Patrick Massot.
\newblock {Formalising Perfectoid Spaces}.
\newblock In Jasmin Blanchette and Catalin Hritcu, editors, {\em Proceedings of
  the 9th {ACM} {SIGPLAN} International Conference on Certified Programs and
  Proofs, {CPP} 2020, New Orleans, LA, USA, January 20-21, 2020}, pages
  299--312. {ACM}, 2020.
\newblock \href {https://doi.org/10.1145/3372885.3373830}
  {\path{doi:10.1145/3372885.3373830}}.

\bibitem{TypeTheory}
Mario Carneiro.
\newblock {\em The {T}ype {T}heory of {L}ean}.
\newblock Springer, Berlin, Heidelberg, 2019.
\newblock {M}aster thesis.
\newblock URL: \url{https://github.com/digama0/lean-type-theory/releases}.

\bibitem{CF}
J.~W.~S. Cassels and A.~Fröhlich (eds.).
\newblock {\em {Algebraic Number Theory}}.
\newblock Academic Press, London; Thompson Book Co., Inc., Washington, D.C.,
  1967.

\bibitem{Lean3}
L.~de~Moura, S.~Kong, J.~Avigad, F.~van Doorn, and J.~von Raumer.
\newblock {The Lean Theorem Prover (System Description)}.
\newblock In Felty A. and Middeldorp A., editors, {\em Automated Deduction -
  CADE-25}, volume 9195 of {\em Lecture Notes in Computer Science}, pages
  378--388. Springer, Cham, 2015.
\newblock \href {https://doi.org/10.1007/978-3-319-21401-6_26}
  {\path{doi:10.1007/978-3-319-21401-6_26}}.

\bibitem{Janusz}
Gerald~J. Janusz.
\newblock {\em {Algebraic Number Fields }}, volume~55 of {\em {Pure and Applied
  Mathematics}}.
\newblock Academic Press, London, 2nd edition, 1996.

\bibitem{Langlands}
R.~P. Langlands.
\newblock {Problems in the Theory of Automorphic Forms}.
\newblock In {\em {Lectures in Modern Analysis and Applications III}}, volume
  170 of {\em {Lecture Notes in Mathematics}}, pages 18--61. Springer, Berlin,
  Heidelberg, 1970.
\newblock \href {https://doi.org/10.1007/BFb0079065}
  {\path{doi:10.1007/BFb0079065}}.

\bibitem{padicsLean}
Robert~Y. Lewis.
\newblock {A Formal Proof of Hensel's Lemma over the p-Adic Integers}.
\newblock In {\em Proceedings of the 8th ACM SIGPLAN International Conference
  on Certified Programs and Proofs}, CPP 2019, page 15–26, New York, NY, USA,
  2019. Association for Computing Machinery.
\newblock \href {https://doi.org/10.1145/3293880.3294089}
  {\path{doi:10.1145/3293880.3294089}}.

\bibitem{Milne}
J.~S. Milne.
\newblock {C}lass {F}ield {T}heory (v4.03), 2020.
\newblock URL: \url{https://www.jmilne.org/math/CourseNotes/CFT.pdf}.

\bibitem{Neukirch99}
J\"{u}rgen Neukirch.
\newblock {\em Algebraic Number Theory}.
\newblock Springer, Berlin, Heidelberg, 1999.
\newblock \href {https://doi.org/10.1007/978-3-662-03983-0}
  {\path{doi:10.1007/978-3-662-03983-0}}.

\bibitem{padicsCoq}
Álvaro Pelayo, Vladimir Voevodsky, and Michael~A. Warren.
\newblock {A univalent formalization of the p-adic numbers}.
\newblock {\em Mathematical Structures in Computer Science}, 25(5):1147–1171,
  2015.
\newblock \href {https://doi.org/10.1017/S0960129514000541}
  {\path{doi:10.1017/S0960129514000541}}.

\bibitem{Stich}
Henning Stichtenoth.
\newblock {\em {Algebraic Function Fields and Codes}}.
\newblock Universitext. Springer, 1993.
\newblock URL: \url{https://dblp.org/rec/books/daglib/0084861.bib}.

\bibitem{HaarLean}
Floris van Doorn.
\newblock {Formalized Haar Measure}.
\newblock In Liron Cohen and Cezary Kaliszyk, editors, {\em 12th International
  Conference on Interactive Theorem Proving (ITP 2021)}, volume 193 of {\em
  Leibniz International Proceedings in Informatics (LIPIcs)}, pages
  18:1--18:17, Dagstuhl, Germany, 2021. Schloss Dagstuhl -- Leibniz-Zentrum
  f{\"u}r Informatik.
\newblock URL: \url{https://drops.dagstuhl.de/opus/volltexte/2021/13913}, \href
  {https://doi.org/10.4230/LIPIcs.ITP.2021.18}
  {\path{doi:10.4230/LIPIcs.ITP.2021.18}}.

\end{thebibliography}

\end{document}